\documentstyle{amsppt}
\newcount\mgnf\newcount\tipi\newcount\tipoformule\newcount\greco 
\tipi=2          
\tipoformule=0   

\global\newcount\numsec\global\newcount\numfor
\global\newcount\numapp\global\newcount\numcap
\global\newcount\numfig\global\newcount\numpag
\global\newcount\numnf
\global\newcount\numtheo

\def\SIA #1,#2,#3 {\senondefinito{#1#2}%
\expandafter\xdef\csname #1#2\endcsname{#3}\else
\write16{???? ma #1,#2 e' gia' stato definito !!!!} \fi}

\def \FU(#1)#2{\SIA fu,#1,#2 }

\def\etichetta(#1){(\veroparagrafo.\veraformula)%
\SIA e,#1,(\veroparagrafo.\veraformula) %
\global\advance\numfor by 1%
\write15{\string\FU (#1){\equ(#1)}}%
\write16{ EQ #1 ==> \equ(#1)  }}

\def\etichettat(#1){\veroparagrafo.\veratheorema:%
\SIA e,#1,{\veroparagrafo.\veratheorema} %
\global\advance\numtheo by 1%
\write15{\string\FU (#1){\thu(#1)}}%
\write16{ TH #1 ==> \thu(#1)  }}

\def\etichettaa(#1){(A\veraappendice.\veraformula)
 \SIA e,#1,(A\veraappendice.\veraformula)
 \global\advance\numfor by 1
 \write15{\string\FU (#1){\equ(#1)}}
 \write16{ EQ #1 ==> \equ(#1) }}
\def\getichetta(#1){Fig. \verafigura
 \SIA g,#1,{\verafigura}
 \global\advance\numfig by 1
 \write15{\string\FU (#1){\graf(#1)}}
 \write16{ Fig. #1 ==> \graf(#1) }}
\def\retichetta(#1){\numpag=\pgn\SIA r,#1,{\verapagina}
 \write15{\string\FU (#1){\rif(#1)}}
 \write16{\rif(#1) ha simbolo  #1  }}
\def\etichettan(#1){(n\verocapitolo.\veranformula)
 \SIA e,#1,(n\verocapitolo.\veranformula)
 \global\advance\numnf by 1
\write16{\equ(#1) <= #1  }}

\newdimen\gwidth
\gdef\profonditastruttura{\dp\strutbox}
\def\senondefinito#1{\expandafter\ifx\csname#1\endcsname\relax}
\def\BOZZA{
\def\alato(##1){
 {\vtop to \profonditastruttura{\baselineskip
 \profonditastruttura\vss
 \rlap{\kern-\hsize\kern-1.2truecm{$\scriptstyle##1$}}}}}
\def\galato(##1){ \gwidth=\hsize \divide\gwidth by 2
 {\vtop to \profonditastruttura{\baselineskip
 \profonditastruttura\vss
 \rlap{\kern-\gwidth\kern-1.2truecm{$\scriptstyle##1$}}}}}
\def\verapagina{
{\romannumeral\number\numcap}.\number\numsec.\number\numpag}}

\def\alato(#1){}
\def\galato(#1){}
\def\veroparagrafo{\number\numsec}\def\veraformula{\number\numfor}
\def\veraappendice{\number\numapp}
\def\verapagina{\number\pageno}\def\veranformula{\number\numnf}
\def\verafigura{{\romannumeral\number\numcap}.\number\numfig}
\def\verocapitolo{\number\numcap}\def\veranformula{\number\numnf}
\def\veratheorema{\number\numtheo}
\def\Eqn(#1){\eqno{\etichettan(#1)\alato(#1)}}
\def\eqn(#1){\etichettan(#1)\alato(#1)}
\def\TH(#1){{\etichettat(#1)\alato(#1)}}
\def\thv(#1){\senondefinito{fu#1}$\clubsuit$#1\else\csname fu#1\endcsname\fi} 
\def\thu(#1){\senondefinito{e#1}\thv(#1)\else\csname e#1\endcsname\fi}

\def\Eq(#1){\eqno{\etichetta(#1)\alato(#1)}}
\def\eq(#1){\etichetta(#1)\alato(#1)}
\def\Eqa(#1){\eqno{\etichettaa(#1)\alato(#1)}}
\def\eqa(#1){\etichettaa(#1)\alato(#1)}
\def\dgraf(#1){\getichetta(#1)\galato(#1)}
\def\drif(#1){\retichetta(#1)}

\def\eqv(#1){\senondefinito{fu#1}$\clubsuit$#1\else\csname fu#1\endcsname\fi}
\def\equ(#1){\senondefinito{e#1}\eqv(#1)\else\csname e#1\endcsname\fi}
\def\graf(#1){\senondefinito{g#1}\eqv(#1)\else\csname g#1\endcsname\fi}
\def\rif(#1){\senondefinito{r#1}\eqv(#1)\else\csname r#1\endcsname\fi}
\def\bib[#1]{[#1]\numpag=\pgn
\write13{\string[#1],\verapagina}}

\def\include#1{
\openin13=#1.aux \ifeof13 \relax \else
\input #1.aux \closein13 \fi}

\openin14=\jobname.aux \ifeof14 \relax \else
\input \jobname.aux \closein14 \fi
\openout15=\jobname.aux
\openout13=\jobname.bib


\ifnum\tipoformule=1\let\Eq=\eqno\def\eq{}\let\Eqa=\eqno\def\eqa{}
\def\equ{}\fi


{\count255=\time\divide\count255 by 60 \xdef\hourmin{\number\count255}
        \multiply\count255 by-60\advance\count255 by\time
   \xdef\hourmin{\hourmin:\ifnum\count255<10 0\fi\the\count255}}

\def\oramin{\hourmin }

\def\data{\number\day/\ifcase\month\or january \or february \or march \or
april \or may \or june \or july \or august \or september
\or october \or november \or december \fi/\number\year;\ \oramin}

\setbox200\hbox{$\scriptscriptstyle \data $}

\newcount\pgn \pgn=1
\def\foglio{\number\numsec:\number\pgn
\global\advance\pgn by 1}
\def\foglioa{A\number\numsec:\number\pgn
\global\advance\pgn by 1}

\footline={\rlap{\hbox{\copy200}}\hss\tenrm\folio\hss}


\global\newcount\numpunt

\magnification=1000
\baselineskip=14pt
\parskip=8pt

\voffset=2.5truepc
\hoffset=0.5truepc
\hsize=6.1truein
\vsize=8.4truein 
{\headline={\ifodd\pageno\rightheadline \else \leftheadline \fi}}
\def\rightheadline{\it  {Metastablity}\hfil\tenrm\folio}
\def\leftheadline{\tenrm \folio \hfil\it  {Bovier et al.}}

\def\d{\delta}
\def\e{\epsilon}

\def\f{\phi}

\def\l{\lambda}

\def\s{\sigma}
\def\t{\tau}

\def\G{\Gamma}

\def\1{{1\kern-.25em\roman{I}}}
\def\eu{{1\kern-.25em\roman{I}}}
\def\f1{{1\kern-.25em\roman{I}}}
\def\N{{\Bbb N}}  
\def\P{{\Bbb P}}  
\def\Q{{\Bbb Q}}  
\def\E{{\Bbb E}}  




\let\cal=\Cal

\def\GG{{\cal G}}

\def\MM{{\cal M}}

\def\chap #1#2{\line{\ch #1\hfill}\numsec=#2\numfor=1\numtheo=1}

\def\ba{{\backslash}}


\def\note#1{\footnote{#1}}

\def\frac#1#2{{#1\over #2}}

\def\text#1{\quad{\hbox{#1}}\quad}

\def\theo #1{\noindent{\thbf Theorem {#1} }}

\def\definition #1{\noindent{\thbf Definition {#1} }}

\def\remark{\noindent{\bf Remark: }}
\def\thanks{\noindent{\bf Acknowledgements: }}

\font\thbf=cmbxsl10 scaled\magstephalf

\font\ch=cmbx12

\font\it=cmti10
\font\bf=cmbx10


\def\vep{\varepsilon}

\font\tit=cmbx12
\font\aut=cmbx12

\centerline{\tit METASTABILITY AND SMALL EIGENVALUES }
\vskip.2truecm
\centerline{\tit IN  MARKOV CHAINS}
\vskip0.5truecm

\centerline{\aut Anton Bovier 
\note{ Weierstrass-Institut f\"ur Angewandte Analysis und Stochastik,
Mohrenstrasse 39, D-10117 Berlin,\hfill\break Germany.
 e-mail: bovier\@wias-berlin.de},
Michael Eckhoff\note{Institut f\"ur Mathematik, Universit\"at Potsdam,
 Am Neuen Palais 10, D-14469 Potsdam, Germany.\hfill\break
e-mail: eckhoff\@rz.uni-potsdam.de},}
\centerline{\aut V\'eronique Gayrard\note{DMA, EPFL, CH-1021 Lausanne, 
Switzerland, and
Centre de Physique Th\'eorique, CNRS,
Luminy, Case 907, F-13288 Marseille, Cedex 9, France. email: Veronique.Gayrard\@epfl.ch},
Markus Klein\note {Institut f\"ur Mathematik, Universit\"at Potsdam,
Am Neuen Palais 10, D-14469 Potsdam, Germany.\hfill\break
e-mail: mklein\@felix.math.uni-potsdam.de}}

\vskip0.5truecm\rm
\def\s{\sigma}
\parskip=0pt
\noindent {\bf Abstract:} 
In this letter we announce rigorous results that elucidate the relation 
between {\it metastable states} and {\it low-lying eigenvalues} in Markov
chains in a much more general setting and with considerable greater precision
as was so far available. 
This includes a {\it sharp} uncertainty principle relating all low-lying 
eigenvalues to mean times of metastable transitions, a relation between the 
support of eigenfunctions and the attractor of a metastable state, and 
sharp estimates on the convergence of 
probability distribution of the metastable transition 
times to the exponential distribution.


\vskip .5cm
\chap{I. Introduction.}1

The phenomenon of {\it metastability} has been a fascinating topic of
non-equilibrium statistical mechanics for a long time. Currently,
it has found renewed interest in the investigation of
{\it glassy systems} and {\it aging phenomena} which appear to play a central 
role in many physical and non-physical systems. An approach to link 
metastability to spectral characteristics, in particular 
{\it low-lying eigenvalues} and the corresponding eigenfunctions has
been proposed by Gaveau and Schulman [GS]. Such an approach is appealing 
not only because it allows to characterize metastability in terms that
are intrinsically dynamic and make no a priori reference to geometric concepts
 such as ``free energy landscapes'', but also since it allows numerical 
investigations of metastable states via numerical spectral analysis (see
Sch\"utte et al. [S,SFHD] for applications to conformational 
dynamics of biomolecules).

Relating metastability to spectral characteristics of the Markov generator
or transition matrix is in fact a rather old topic. First mathematical 
results go back at least as far as Wentzell [W] (see also [M] for 
more recent results) and Freidlin and
Wentzell (see [FW]). Freidlin and Wentzell relate the eigenvalues 
of the transition matrix for Markov processes with exponentially small
transition probabilities to exit times from ``cycles''; Wentzell 
has a similar result for the spectral gap in the case of certain diffusion 
processes. All these relations are on the level of logarithmic equivalence,
i.e. of the form $\lim_{\e\downarrow 0} \e \ln (\l^\e_i T^\e_i) =0$
where $\e$ is the small parameter, and $\l^\e_i, T^\e_i$ are
the eigenvalues, resp. exit times. This rather crude level of precision 
persists also in the more recent literature and prevents, in particular, 
applications to systems with unbounded numbers of metastable states
which are particularly relevant for glassy systems.

In this letter we announce results that -- for a large class of Markov
chains -- improve this situation considerably: in particular we allow
for the number of metastable states to grow (with e.g. the `volume'),
and we give precise control of error terms for `finite volume' systems.
Moreover, we provide representations for all quantities concerned that
are computable in terms of certain `escape probabilities' that are in turn 
well controllable through variational representations [BEGK1]. A more detailed
exposition of our results, as well as the proofs, will be given in 
two forthcoming papers [BEGK2,EK].
\medskip
\chap{2. Stable set and metastable states.}2
\medskip
We will consider in the sequel Markov chains $X_t$ with state space $\G_N$, 
discrete time\note{All results apply, however,
also to  continuous time.} $t\in\N$, and transition matrices $P_N$. We will 
assume that for any fixed $N$ they are ergodic, and have a unique invariant 
distribution $\Q_N$. We are interested in the situation when these
chains exhibit ``metastable'' behaviour; loosely speaking, this means that the 
state space $G_N$ can be decomposed into subsets $S_{N,i}$ such that the
 typical times the process takes to go from one such set to another are
much larger than the time it takes to ``look like'' being in equilibrium
with respect to the conditional distribution $\Q_N(\cdot|S_{N,i})$. Some 
reflection shows that this statement has considerable difficulties and
cannot be interpreted literally, and that a precise definition of 
metastability is a rather tricky business (see e.g. the recent discussion in 
[BK]). We will give a precise definition that is, however, inspired by this
vague consideration. The main point here is that one should make precise the
two time scales we alluded to. We will take the following attitude: 
to look ergodic within $S_{N,i}$, the process should have at least 
enough time to reach the ``most attractive'' state within $S_{N,i}$, while
at least the times to go from two such states in different 
metastable regions should be long compared to that time. Note that this
concept is rather flexible and allows, in general,
to define metastable states corresponding to different time scales.

The following definition of ``stable sets'' follows this 
ideology; however, we prefer to use certain probabilities 
rather than actual times as criteria, mainly because these are
more readily computable. Linking them in a precise manner to times will
be part of our results. We will write $\t^x_I$, for $x\in\G_N$, 
$I\subset\G_N$,
 for the 
first non-zero time the process started at $x$ arrives at $I$.

 \definition {\TH(D.1)} {\it A set $\MM_N\subset\G_N$ 
will be called a set of {\rm stable
points}  if it satisfies the following assumptions: 
There exist finite positive 
constants $a_N$, $b_N$, $c_N$, and  $r_N$ such that
they satisfy  for some sequence $\vep_N$ s.t.
$|\G_N|\vep_N\downarrow 0$,
$a_N^{-1}\leq \vep_N b_N$.
\item{(i)} For all $z\in\G_N$, 
$$
\P\left[\t^z_{\MM_N}<\t^z_z\right] \geq b_N
\Eq(0.2)
$$
\item{(ii)} For any $x\neq y\in\MM_N$, 
$$
\P\left[\t^x_y<\t^x_x\right]\leq a^{-1}_n
\Eq(0.3)
$$
We associate with each $x\in\MM_N$ its {\it local valley}
$$
A(x)\equiv \left\{z\in\G_N: \P\left[\t^z_x=\t^z_{\MM_N}\right]
=   \sup_{y\in\MM_N     } \P\left[\t^z_y=\t^z_{\MM_N}\right]\right\}
\Eq(0.5)
$$
Then 
$$
r_N\geq\frac{\Q_N(x)}{\Q_N(A(x))}\equiv R_x\geq c^{-1}_N
\Eq(0.7)
$$
}
We will also write $T_{x,I}\equiv \P[\t^x_I\leq\t^x_x]^{-1}$.
An important characteristic of the sets $I\subset\MM_N$ is 
$T_I\equiv \sup_{x\in\MM_N} T_{x,I}$.   
A simplifying assumption, that will be seen to ensure sufficient ``spacing''
of the low lying eigenvalues is that of ``genericity'', defined as
follows:

\definition {\TH(D.2)}{\it We say that our Markov chain is {\rm
generic} on the level of the set $\MM_N$, if there exists a sequence 
$\e_N\downarrow 0$, s.t. 
\item{(i)} For all pairs $x,y\in\MM_N$, and any set $I\subset\MM_N\ba\{x,y\}$
either
$T_{x,I}\leq \e_N T_{y,I}$ or
$ T_{y,I}\leq \e_N T_{x,I}$.
}

Each of the elements of $\MM_N$ in the generic case will then correspond
indeed to a metastable state. Our first task is to identify precisely 
the notion of the exit time from a metastable state. To do so, we 
define for any $x\in\MM_N$ the set
$\MM_N(x)\equiv \{y\in\MM_N:\Q_N(y)>\Q_N(x)\}$; these are the 
points that are even more stable than $x$. The metastable exit time, $t_x$
from $x$ is then defined as the time of the first arrival from $x$ in 
$\MM_N(x)$. With this notion we can formulate our main result:

\theo{\TH(A.1)}{\it Assume that $\MM_N$ is a stable set and that the
genericity assumptions are satisfied with $\vep_N$ such that 
$r_Nc_N \e_N\downarrow 0 $. Set $t_x\equiv \t^x_{\MM_N(x)}$. Then
\item{(i)} For any $x\in\MM_N$,
$$
\E \t_x =R^{-1}_xT_{x,\MM_N(x)} (1+ o(1))
\Eq(T.1)
$$
\item{(ii)} For any $x\in\MM_N$, there exists an eigenvalue $\l_x$ 
of $1-P_N$ that satisfies 
$$
\l_x=\frac 1{\E t_x}\left(1+o(1)\right)
\Eq(T.2)
$$
Moreover, there exists a constant $c>0$ such that for all $N$
$$
\s(1-P_N) \ba \cup_{x\in \MM_N} \l_x \subset (c b_N,1]
\Eq(T.3)
$$
\item{(iii)} For any $x\in\MM_N$, for all $t>0$,
$$
\P[t_x >t\E t_x]= e^{-t(1+o(1))}(1+o(1))
\Eq(T.4)
$$
\item{(iv)} If $\psi_x$ denotes the eigenvector of $1-P_N$ corresponding to 
the eigenvalue $\l_x$, then 
$$
\psi_x(y)=\cases \P[\t^y_x<\t^y_{\MM_N(x)}](1+o(1)),&\text {if}  
\P[\t^y_x<\t^y_{\MM_N(x)}] \geq \e_N\cr
O(\e_N),&\text{otherwise}
\endcases\Eq(T.5)
$$
}
\remark Explicit bounds on the error terms are given [BEGK2,EK]. 

Let us make some additional comments on this theorem. First of all,
they identification of what constitutes a metastable exit is crucial, 
and, in particular the fact that these processes include the 
{\it transition through the ``saddle point''}, guaranteed in our case 
by the insistence that the process has actually arrived in $\MM_N(x)$. 
Without taking this into account, the precise uncertainty principle 
(ii) could not hold. It is interesting to note that on the level of this 
theorem, the difficulties associated with the control of the passage through 
a saddle are not visible, and that we have the exact formula \eqv(T.1)
for the mean exit time. Of course the difficulty is hidden in the 
quantities $T_{x,y}$ whose computation is far from trivial. However, we
have shown in [BEGK1] that at least in the reversible case, using 
a variational representation, very precise control of such quantities can 
be gained in certain setting. Somewhat less precise results can also be 
obtained in some non-reversible situations [EK]. 
Concerning our estimate on the eigenfunction, it is easy to see that [BEGK2]
$ \P[\t^y_x<\t^y_{\MM_N(x)}]$ is either very close to one or very close to 
zero, except on a set of points whose invariant measure is extremely small.
Therefore, the corresponding right-eigenfunctions 
$\psi^r_x(z)=\Q_N(z)\psi_x(z)$, are essentially proportional
to the measure $\Q_N$ conditioned on the local valley of corresponding to $x$
(all up to errors of order $\e_N$), i.e. they do indeed represent 
{\it metastable measures}, as suggested  in [GS].  

\medskip
\chap{3. Some ideas of the proofs.}3
\medskip
The first major ingredient for the proof is a representation formula
for the Green's function of the transition matrix $P_N$  in terms of
certain probabilities. It implies in particular that for any $I\subset \G_N$,
$$
\E t^x_I=\sum_{y\in\G_N\ba I\ba x} \frac{\Q_N(y)}{\Q_N(x)}{\P[\t^y_x<\t^y_I]}
{\P[\t^x_I<\t^x_x]}+\frac 1{\P[\t^x_I<\t^x_x]}
\Eq(3.1)
$$
This formula was first derived for the reversible case in [BEGK1]. 
An apparently independent derivation that also covers the non-reversible 
case was given recently in [GM]. This formula allows to prove 
in a rather simple way, \eqv(T.1). However, the realization that this formula
actually arises from a representation of the Greens function makes it
even more useful. 

Our analysis of the spectrum of $1-P_N$ passes through the analysis of the 
Laplace transforms, $G_{y,I}^x(u)\equiv \E e^{ut^x_y}\1_{\t^x_y<\t^x_I}$  
of transition times of process that is `killed' upon 
arrival in a set $I\subset \G_N$. We write $P_N^J$ for the transition 
matrix of such a process, and we write $\l_J$ for the 
smallest eigenvalue of $(1-P^J)$. It then turns out that all 
eigenvalues of $(1-P_N)$ below $\l_{J}$ can be characterised as follows:
Set $u(\l)\equiv -\ln (1-\l)$. Define the  $|J|\times|J|$ 
matrix $\GG_J(u)$ whose elements are
$$
\d_{m',m}-G^{m'}_{m,J}(u),\quad m',m\in J
\Eq(3.2)
$$
Then $\l$ is an eigenvalue of $(1-P_N)$ below $\l_J$, if and only if
$$
\det G_J(u(\l))=0
\Eq(3.3)
$$
This equation is rather easy to understand if $|J|=1$. In this case,
\eqv(3.3) becomes simply
$G^m_m(u(\l)=1$. By a simple renewal argument, one sees that
$
G^m_x(u)=\frac{G^m_{x,m}(u)}{1-G^m_{m,x}(u)}$. Therefore, $u(\l)$ defined by 
\eqv(3.3) is the first value at which $\sup_{x\in G_N} G^m_x(u)=+\infty$.
The general formula \eqv(3.2) is somewhat less intuitive. Basically,
one makes an ansatz for the eigenfunctions of $(1-P_N)$ in terms of the 
Laplace transforms of the form
$$
\Psi(x)=\sum_{m\in J} \phi_m G^x_{m,J}(u)
\Eq(3.4)
$$
One then finds that condition \eqv(3.3) is sufficient 
for the ansatz to yield eigenfunctions wit $u=u(\l)$. Moreover,
one can show that if $\l$ is an eigenvalue, then the eigenfunctions
can be represented in this form and \eqv(3.3) must be satisfied. 

To complete the proof one needs good control over the Laplace transforms; this 
is partly provided again by the representation of the Green's function, 
complemented by lower  bounds on eigenvalues $\l_J$ obtained from a 
Donsker-Varadhan [DV] argument. The actual proofs are rather involved and must be
left to 
the longer publications [BEGK2,EK]. Let us finally mention that 
the good control over the spectrum of $(1-P_N)$ allows in turn a very good 
control of the analytic properties of the Laplace transforms which allow
in turn the sharp estimates on the probability distribution of metastable
transition times stated under (iv). 

\medskip
\chap{References}0
\medskip

\item{[BEGK1]} A. Bovier, M.Eckhoff, V. Gayrard and M. Klein, 
``Metastability in Stochastic 
Dynamics of Disordered Mean-Field Models``, WIAS-preprint 452, to appear in 
Prob. Theor. Rel, Fields, (2000).
\item{[BEGK2]} A. Bovier, M.Eckhoff, V. Gayrard and M. Klein, ``Metastability 
and low-lying spectra in reversible Markov chains``, in perparation (2000).
\item{[BK]} G. Biroli and J. Kurchan, ``Metastable states in glassy systems'',
\hfill\break{\tt http://www.xxx.lanl.gov/cond-mat/0005499} (2000).
\item{[DV]}M.D.  Donsker and S.R.S. Varadhan,  ``On the principal eigenvalue 
of second-order elliptic differential
operators'', 
Comm. Pure Appl. Math. {\bf 29},  595-621 (1976).  
\item{[EK]} M. Eckhoff, M. Klein, ``Metastability and low lying spectra
in non-reversible Markov chains'', in preparation (2000).
\item{[FW]} M.I. Freidlin and A.D. Wentzell, ``Random perturbations of 
dynamical systems'', Springer, Berlin-Heidelberg-New York, 1984.
\item{[GM]} B. Gaveau and M. Moreau, ``Metastable relaxation times and absorbtion
probabilities for multidimensional stochastic systems'', J. Phys. A: Math. Gen.
{\bf 33}, 4837-4850 (2000).
\item{[GS]}  B. Gaveau and L.S. Schulman, ``Theory of nonequilibrium 
first-order phase transitions for stochastic dynamics'',  J.
Math. Phys. {\bf 39}, 1517-1533 (1998).
\item{[M]} P. Mathieu, ``Spectra, exit times and long times asymptotics in 
the zero white noise limit'', Stoch. Stoch. Rep. {\bf 55}, 1-20 (1995).
\item{[S]} Ch. Sch\"utte, ``Conformational dynamics: modelling, theory, 
algorithm, and application to bio\-mole\-cules'', preprint SC 99-18, 
ZIB-Berlin (1999).
\item{[SFHD]} Ch. Sch\"utte, A. Fischer, W.  Huisinga, P. Deuflhard, ``A 
direct approach to conformational dynamics based on
hybrid Monte Carlo'', J. Comput. Phys. {\bf 151}, 146-168 (1999).
\item{[Sc]} E. Scoppola, ``Renormalization and graph methods for Markov 
chains''. Advances in dynamical systems and quantum
physics (Capri, 1993), 260-281, World Sci. Publishing, River Edge, NJ, 1995.
\item{[W]}  A.D. Wentzell, ``On the asymptotic behaviour of the greatest
 eigenvalue of a second order elliptic differential operator with a small 
parameter  in the higher derivatives'', Soviet Math. Docl. {\bf 13}, 13-17 
(1972).

\end